# Formation and Properties of Selenium Double-Helices inside Double-Wall Carbon Nanotubes: Experiment and Theory

*Toshihiko Fujimori,[1] Renato Batista dos Santos,[2,3] Takuya Hayashi,[4] Morinobu Endo,[1] Katsumi Kaneko,[1] and David Tománek[3,\*]*

[1]Research Center for Exotic Nanocarbons (JST), Shinshu University, 4-17-1 Wakasato, Nagano-city 380-8553, Japan

[2]Universidade Federal da Bahia, Instituto de Física, Salvador, Bahia 40.210-340, Brazil

[3]Physics and Astronomy Department, Michigan State University, East Lansing, Michigan 48824, USA

[4]Faculty of Engineering, Shinshu University, 4-17-1 Wakasato, Nagano-city 380-8553, Japan

*Address correspondence to tomanek@pa.msu.edu

**ABSTRACT**

We report the production of covalently bonded selenium double-helices within the narrow cavity inside double-wall carbon nanotubes. The double-helix structure, characterized by high-resolution transmission electron microscopy and X-ray diffraction, is completely different from the bulk atomic arrangement and may be considered a new structural phase of Se. Supporting *ab initio* calculations indicate that the observed encapsulated Se double-helices are radially compressed and have formed from free Se atoms or short chains contained inside carbon nanotubes. The calculated electronic structure of Se double-helices is very different from the bulk system, indicating the possibility to develop a new branch of Se chemistry.

**(101 words/250 words)**

A key objective of Chemistry is to rearrange the bonding configuration of atoms in order to arrive at substances with vastly different properties. Elemental Selenium, in its trigonal bulk structure depicted in Figure 1(a), is a semiconductor known for its photovoltaic properties and high photoconductivity.[1] On the other hand, it is well established that a pressure-induced structural phase transition from the trigonal to a monoclinic phase with a puckered structure[2,3] changes Se from a semiconductor to a metal. The photoconductive properties of bulk Se, which are a consequence of its unique trigonal ($P3_121$) bulk structure[4,5] consisting of three-fold helical chains and shown in Figure 1(a), have been exploited in applications ranging from solar cells, light-switching devices and rectifiers to xerographic photoreceptors and photographic exposure meters.[1] It is likely that other structural arrangements, which could be achieved under different steric constraints, may give rise to yet unexplored properties.

The known properties of the trigonal bulk structure have inspired many experimental[6-8] and theoretical[9] studies of nanowires and nanotubes consisting of trigonal Se. The unusual behavior of Se includes its ability to change the valence configuration to form intriguing systems, such as $Bi_2Se_3$ nanoribbons that show topological insulator behavior,[10-12] CdSe nanocrystals forming quantum dots,[13] and cubane clusters.[14] Since all these Se-based systems are nanometer-sized, we conclude that the ability to Se to change its valence electron configuration in nanostructures offers a rich potential to develop a new branch of Se chemistry in nanometer-sized cavities found inside nanotubes and related systems.

In this paper, we report the production of covalently bonded selenium double-helices within the narrow cavity inside double-wall carbon nanotubes (DWCNTs). The Se double-helix structure inside a DWCNT (Se@DWCNT), which we characterize by high-resolution transmission electron microscopy (HR-TEM) and X-ray diffraction (XRD), is completely different from the atomic arrangement in bulk Se and may be considered a new structural phase. Supporting *ab initio* calculations indicate that the observed encapsulated Se double-helices are radially compressed and have formed from free Se atoms or short chains in the narrow cavity inside the carbon nanotubes. The calculated electronic structure of Se double-helices is very different from the bulk system, suggesting a very different behavior of these constrained nanostructures.

Whereas Se in the bulk trigonal phase, depicted in Figure 1(a), contains covalently connected single-helices with a pitch length of 0.496 nm, the ≈1.9 nm pitch length of the Se@DWCNT double-helix structure, shown in the HR-TEM image in Figure 1(b), is much longer. Essential for the formation of this structure is the extremely narrow confining space inside a quasi-1D carbon nanotube (CNT) with a typical diameter ≈1 nm. The nanocavity inside CNTs, in particular DWCNTs with a small internal diameter, is known to preserve unusual metastable structural arrangements not found in free space. Capillary filling of this nanocavity with metals has been reported soon after the synthesis of single-wall carbon nanotubes in presence of a metal catalyst.[15,16] Even though the atomic structure could not be resolved in the early studies, lattice fringes seen in HR-TEM images indicated crystallinity of enclosed Pb and Ni metals or their compounds.[15,16] The "nanocapillary" force, which pulls in and stabilizes encapsulated atoms and molecules in the confining nanotube volume under ambient pressure, has the same effect as a high-pressure environment. Thus, encapsulation inside a nanotube may completely change the atomic structure of ionic crystals,[17] metal nanowires,[18,19] ordering of water molecules,[20,21] and even convert hydrocarbons to diamond nanowires.[22] Whereas iodine has been known to form helical chains inside CNTs,[23,24] no such microscopic structural information has been provided in former reports[25-29] of Se contained in the narrow space inside CNTs.

**RESULTS AND DISCUSSION**

Se@DWCNT samples used in our studies were produced by exposing open-ended DWCNTs to sublimed elemental Se, as described in the Methods section. Representative HR-TEM images of an empty and a Se filled DWCNT are shown in Figures 2(a) and 2(b), respectively. Clearly visible in Figure 2(b) is the atomically resolved Se structure contained inside the inner CNT, characterized by a double-helix arrangement with a pitch length of ≈2nm. This structure is significantly different from the single-helix arrangement in bulk Se, shown in Figure 1(a), with the much smaller pitch length of 0.496 nm.

To better understand the double-helix arrangement, we subjected the TEM image in Figure 2(b) to a fast Fourier transformation (FFT), yielding the top panel of Figure 2(c). This figure was subsequently subjected to an inverse FFT treatment, yielding the lower panel of Figure 2(c). The image in the lower panel of Figure 2(c), representing the direct structure, shows a periodically repeating pattern of single and double lines inside the inner CNT, which we associate with Se chains. Figure 2(d) shows line intensity profiles within the image in the lower panel of Figure 2(c) in the direction normal to the tube axis, indicated by the horizontal red arrow. Noticeable differences between the intensity profiles taken at positions (i)-(iv), indicated in Figure 2(c), provide a quantitative characterization of the double-helix structure as well as pitch length and helix diameter.

During TEM observations, we occasionally noticed a structural transformation from a Se double-helix to a single-helix structure, as seen in Figure 3, apparently induced by the electron-beam irradiation during the observation. Figure 3(a) shows a Se double helix inside the inner wall of a DWCNT at the initial stage of the TEM observation. Subsequently, the atomic arrangement within the Se double-helix changes in a manner, which can not be easily identified, as seen in Figure 3(b). Next, the upper part of the Se structure transforms to a single helix, shown in Figure 3(c), with the observed helical pitch length of 0.45-0.50 nm, which is close to the bulk Se value,[5] 0.496 nm. Finally, the single helix detaches from the double helix structure and disappears from the observed area, as evidenced in Figure 3(d). We believe that the structural change was not affected by the proximity to the DWNT end, but was induced by enhancing the electron-beam exposure in a restricted region of the image.

To confirm that the periodicity observed in the HR-TEM image is representative of the entire Se@DWCNT system, we used X-ray diffraction for a quantitative structural characterization. Figure 4(a) displays the XRD profiles of Se@DWCNTs measured at $T$=300-600 K and compares them to empty DWCNTs. The XRD profile of Se@DWCNTs ($T$=300 K) shows a broad Bragg peak at $2\theta$=2.45°, indicating a periodic structure with the period $d\approx$1.87 nm, very close to our HR-TEM observations in Figure 1(b). This period is significantly larger than the 0.438 nm separation between adjacent Se chains in bulk Se and the ≈0.2 nm separation between the Se strands in the double-helix structure in the HR-TEM image. Intriguingly, the axial pitch length $d_{axis}$ of Se double-helices decreases significantly from 1.87 nm at $T$=300 K to 1.76 nm at the higher temperature of $T$=450 K, as seen in Figure 4(b). This thermal shrinkage is more pronounced in Se double helices, but agrees with the trend of $d_{c//}$ along the $c$ axis that has been reported for bulk Se and is shown near the bottom of Figure 4(b). Above $T$=450 K, this trend reverses for Se@DWCNTs and the axial pitch length $d_{axis}$ starts increasing again with increasing temperature. Most important, we can observe the Bragg peak corresponding to the period of the double-helices even at significantly higher temperatures than the melting point $T_{m.p.}$=490 K of bulk trigonal Se,[1] confirming an unusually high thermal stability of the Se double-helix phase inside the DWCNT cavity.

To find out whether free-standing single and double helices of Se are stable, we performed extensive structure optimization calculations for these systems and present the optimum structure of a Se single-helix in Figure 5(a) and of a double-helix in Figure 5(b). For the sake of comparison, we superposed in these figures the optimum Se helix structure with that of a (5,5) carbon nanotube, which we will consider later on. Additional information about the structure and simulated TEM images of these helices is provided in Figures S1-S4 in the supporting information. We found good agreement between the calculated and observed structures not only for uniform single and double helices, but also double-helices connecting to a single-helix in Figure S5 and to a ladder structure in Figure S6 of the supporting information.

The equilibrium structure of Se helices encapsulated inside a CNT is likely to differ from free-standing Se helices. Precise atomistic calculations of the equilibrium structure and the interaction of encapsulated Se helices with the surrounding DWCNT are not practical, as they would require extremely large unit cells containing thousands of carbon atoms in the surrounding nanotube. We used a much more practical approach that addresses all essential parts of the system and provides an easy interpretation of the results. The basic assumption is that the atomic structure of the nanotube surrounding the Se helix plays only a secondary role and that the nanotube may be considered a rigid container with cylindrical symmetry. As discussed in more detail in Figure S7 of the supporting information, we determined the interaction of a Se helix with the surrounding DWCNT by approximating the nanotube wall by a graphene monolayer that interacts with a Se chain. We found that the total energy of the system as a function of the Se-graphene separation can be approximated well by a smooth interaction potential. The potential has a shallow minimum at 0.32 nm, corresponding to a binding energy of 0.2 eV per Se atom, and is repulsive at small Se-C separations. We used this potential with cylindrical symmetry to estimate the effect of the DWCNT on the equilibrium geometry and the vibrational spectrum of the enclosed selenium structure. Vibrational spectra of Se helices in nanotubes of different diameter, modeled by a constraining potential, are presented in Figures S8-S9 of the supporting information. We found that vibrations of confined helices resemble those of free Se helices in wide nanotubes with a diameter exceeding 0.84 nm, but change substantially inside narrower nanotubes.

Considering the structure of a Se single-helix with radius $R_H$=0.09 nm according to Figure 5(a), we should expect an optimum fit inside a nanotube with the diameter 2×(0.09+0.32) nm = 0.82 nm. This, as a matter of fact, is the inner diameter of the DWCNTs used in our experimental study. The structure of Se inside wider nanotubes should be very similar to free-standing Se helices shown in Figure 5. In narrower nanotubes, the main effect of the surrounding wall is to compress the Se helices radially, thereby increasing the helical pitch, as illustrated in Figure S4 of the supporting information. We found that as the diameter of the enclosing model nanotube was reduced from 1.0 nm to 0.6 nm, the pitch of the Se single-helix increased to 0.7 nm and the helix resembled a linear chain at an energy cost of 0.52 eV per Se atom. Since this deformation energy investment exceeds significantly the energy gain of 0.2 eV/Se atom due to the attractive Se-CNT interaction, we should not expect spontaneous encapsulation of Se helices in nanotubes with an inner diameter much below 0.8 nm. In absence of any other constraints, we should observe only small deformation of Se structures with respect to the free helical geometries.

The Se double-helix structures reported in this study have a much smaller diameter than expected based on calculations. Since the radial compression of the double-helix inside DWNTs with an inner diameter of 0.8 nm requires more energy than the expected gain upon encapsulation, we conclude that the double helices have not been created outside the DWCNTs. A much more likely scenario involves encapsulation of Se atoms or short chains in the narrow space inside DWCNTs, where they reconnect to a double-helix structure. The equilibrium shape of the Se double-helix is given by an energetic compromise between minimizing the Se-DWCNT repulsion and the radial compression energy of the double-helix. We found that such a radial compression causing a reduction of the double-helix radius to 0.1 nm increased the pitch length beyond 1.0 nm. At nonzero temperatures, where Se double-helices may uncoil at little energy cost, the pitch length estimated from calculations agrees with the observed value between 1-2 nm.

To better understand the electronic properties of systems containing Se and carbon nanotubes, we calculated the electronic structure of a Se single- and double-helix, a narrow (5,5) carbon nanotube and a Se single-helix enclosed inside this nanotube. Our results for the band structure and electronic density of states (DOS) of these systems are shown in Figure 6. The electronic spectrum of a free-standing Se single-helix, shown in Figure 6(a), is dominated by van Hove singularities at the band edges and displays a fundamental band gap of 1.6 eV. Since the band gap is likely underestimated in DFT calculations, we expect the Se single-helix to be a semiconductor with an even larger band gap. The corresponding results for a free-standing Se double-helix with a much wider diameter, shown in Figure 6(b), are significantly different. For one, the fundamental band gap decreased significantly to 0.1 eV. The van Hove singularities do not dominate the DOS as in a single-helix, since Se atoms hybridize not only with the two neighbors along the same helical strand, but also with neighboring Se atoms in the second helical strand. Even a small hybridization with the surrounding nanotube will turn the Se@DWCNT system metallic. We find the calculated DOS in the valence band region of free-standing Se helices consistent with the observed X-ray photoelectron spectra (XPS) spectra presented in Figure S10 of the supporting information. The spectrum of a free-standing (5,5) nanotube in Figure 6(c) is shown mainly to better understand the new features in the spectrum introduced by the encapsulation of a Se helix, as seen in Figure 6(d). We chose the narrow (5,5) nanotube as a computationally manageable model system that describes the essentials, but does not necessarily reproduce all features of the complex systems observed in the experiment. Since the diameter of the (5,5) nanotube is smaller than that of the nanotubes in our experiment, we considered a single and not a double-helix of Se inside this nanotube. The calculated DOS in Figure 6(d) resembles a superposition of the densities of states of the Se helix and the CNT, reflecting only a small Se-C hybridization, consistent with the predicted small Se-C bonding. The presence of a selenium-derived van Hove singularity near the Fermi level suggests the possibility to strongly enhance conductivity of this system by doping. The small calculated C-Se interaction is responsible for nearly negligible C*1s* core-level shifts in the XPS spectra of pristine DWCNTs and Se@DWCNTs, shown in Figure S11 of the supporting information.

To obtain a different view of the interaction between the Se helix and the surrounding carbon nanotube, we studied the charge redistribution in this system. Our results for the electron density difference $\Delta n(\mathbf{r})$ between the Se@CNT system and a superposition of neutral atoms, shown in Figure 7, indicate a small electron transfer from Se to the C atoms of the surrounding nanotube. A very small net positive charge 0.17 e on the Se atoms, obtained from a Mulliken population analysis with a single-$\zeta$ basis and discussed in Figure S12 of the supplementary information, is consistent with the fact that C and Se share the identical Pauling electronegativity value of 2.55. The $\Delta n(\mathbf{r})$ contour plot in Figure 7 shows no regions of strong electron accumulation, providing additional support for our conclusion that the electronic interaction between the selenium and the carbon system is very small.

**CONCLUSIONS**

In conclusion, we report the production of a new crystalline phase of selenium within the narrow cavity inside DWCNTs by exposing open-ended nanotubes to Se vapor. The atomic arrangement, characterized by covalently bonded Se double-helices, is completely different from that in bulk Se. We characterized the structure by high-resolution transmission electron microscopy, X-ray diffraction, and core-level spectroscopy. Supporting *ab initio* calculations indicate that the observed encapsulated Se double helices are radially compressed and have formed from free Se atoms or short chains in the narrow cavity inside the carbon nanotubes. The calculated electronic structure of Se double helices differs significantly from that of bulk Se, indicating unusual conductance behavior and the possibility to develop a new branch of Se chemistry.

**EXPERIMENTAL METHODS**

We used highly crystalline DWCNTs, purchased from TORAY Industries, Inc., with inner tube diameter of ≈0.8 nm. In order to remove the terminating end-caps and insert Se into the nanocavity, the DWCNTs were first oxidized at 723 K under dry air (100 ml min-1) for 1 h. The oxidized DWCNTs and Se pellets (99.99%, Wako Pure Chemical Industries, Ltd.) were sealed in a forked glass tube *in vacuo* (<1 Pa) and subsequently kept at 973 K for 48 h. The as-prepared sample was then washed with carbon disulfide under ultrasonication for 5 min, and then the DWCNT-dispersed in carbon disulfide solution was filtered for extracting the DWCNT sample. This process was repeated three times. The filtered sample was then heated at 423 K under vacuum (<1 Pa) to remove excess Se attached to the outside of the DWCNTs. DWCNTs containing encapsulated Se (Se@DWCNTs) were then characterized by a double Cs-corrected (CEOS GmbH) HR-TEM (JEM-2100F, JEOL) operated at 80 kV. The samples were further characterized by X-ray diffraction (XRD) with $\lambda$=0.08003 nm at the synchrotron source SPring-8 and X-ray photoelectron spectroscopy (XPS) using a monochromatised AlKα X-ray source under $10^{-6}$ Pa (AXIS-ULTRA DLD, Shimazu).

**THEORETICAL METHODS**

Our calculations of the equilibrium structure, stability and electronic properties of Se chains inside CNTs have been performed using *ab initio* DFT as implemented in the SIESTA code.[30] We used the Ceperley-Alder[31] exchange-correlation functional as parameterized by Perdew and Zunger,[32] norm-conserving Troullier-Martins pseudopotentials,[33] and a double-ζ basis including polarization orbitals. We used periodic boundary conditions to represent arrays of aligned, but well separated CNTs containing Se or arrays of separated Se chains on graphene. The Brillouin zone of the isolated 1D chain and nanotube structures was sampled by 10 *k*-points and that of chains interacting with a graphene monolayer by a fine $10\times10$ k-point grid.[34] We used a mesh cutoff energy of 100 Ry to determine the self-consistent charge density, which provided us with a precision in total energy of <1 meV/atom.

**FIGURES**

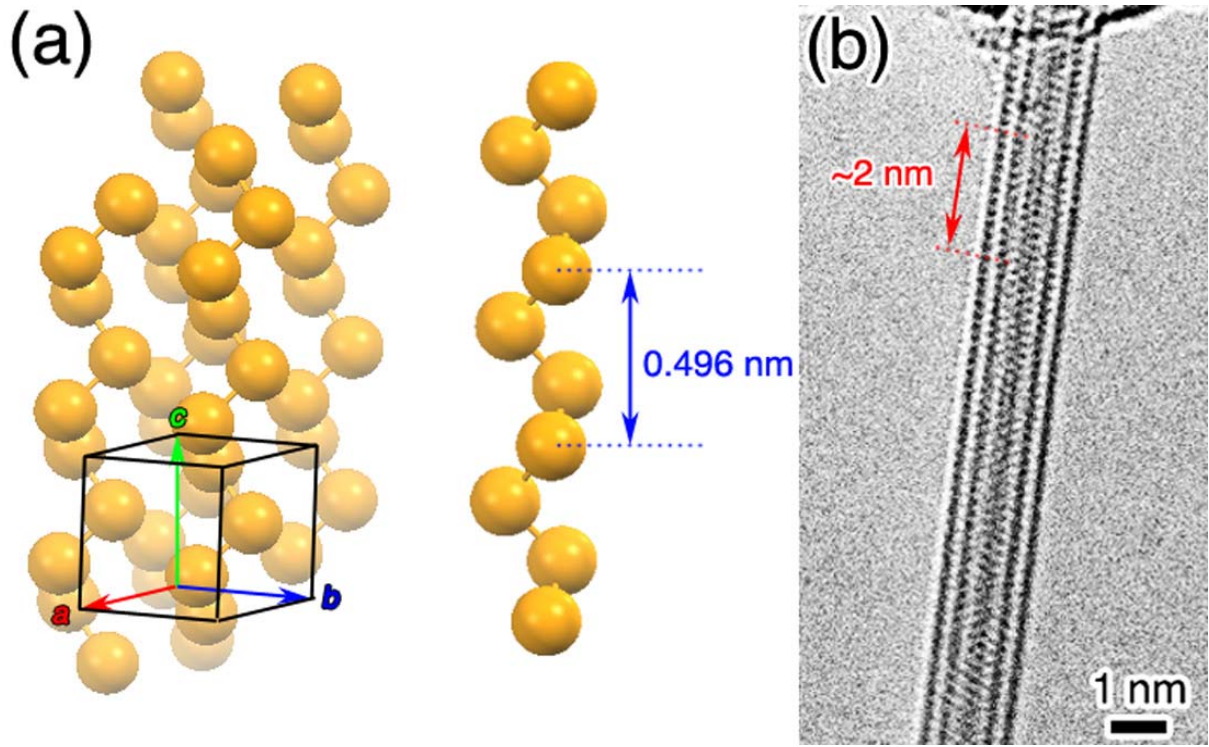

**Figure 1.** (a) Crystallographic structure of trigonal Se (P3$_1$21) consisting of four Se chains in a unit cell (left) and the constituent single Se chain (right) with a pitch length of 0.496 nm.[5] (b) HR-TEM image of Se@DWCNT, revealing the Se double-helix structure with a pitch length of ~2 nm.

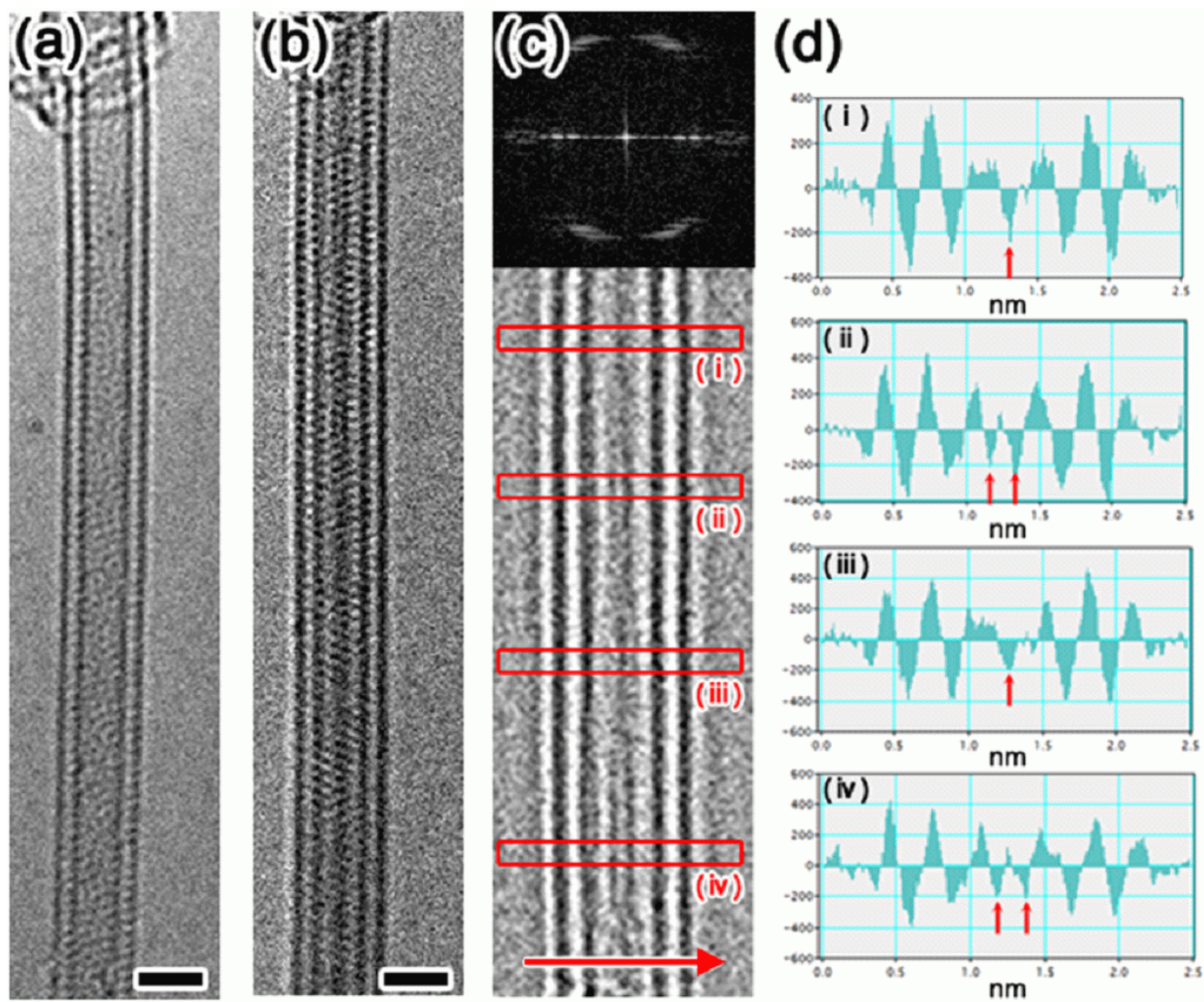

**Figure 2.** HR-TEM images of (a) an empty DWCNT and (b) Se@DWCNT. Scale bar, 1 nm. The Fast Fourier Transform (FFT) of image (b) is presented in the upper panel of (c). The lower panel of (c) contains the inverse FFT of the upper panel. (d) Line intensity profiles of the image in the lower panel of (c), shown along the direction of the red horizontal arrow at positions (i)-(iv). Red vertical arrows in (d) highlight the positions, where single or double lines associated with Se presence occur in the lower panel of (c).

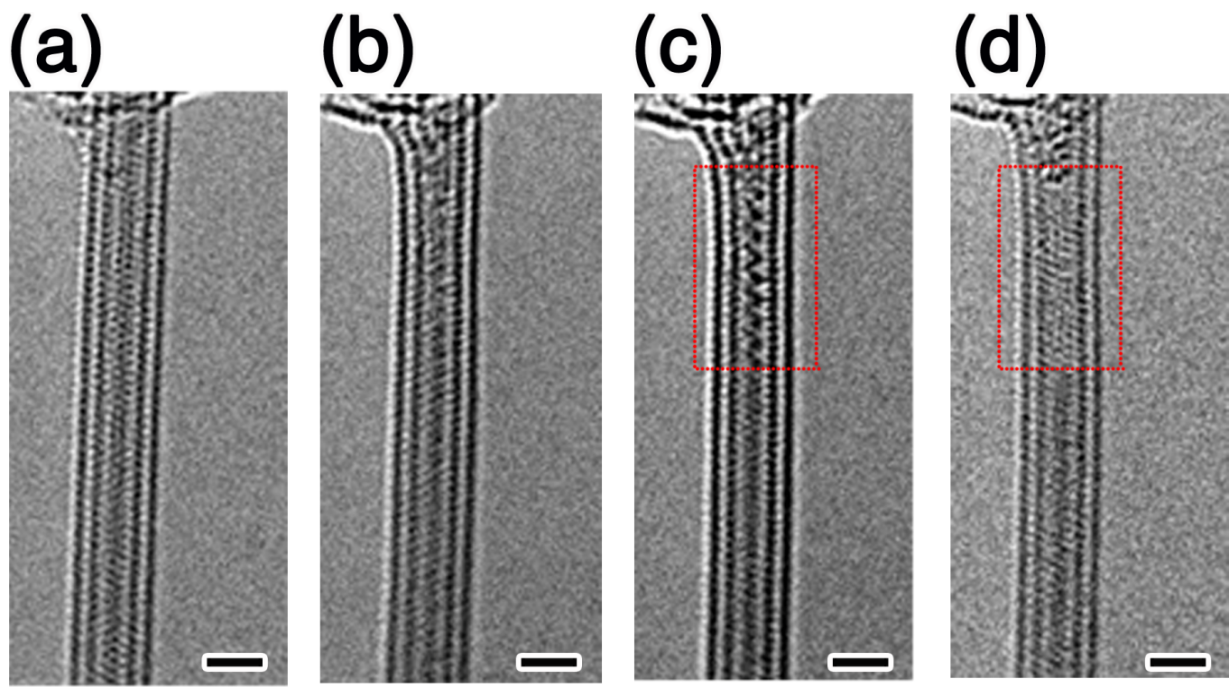

**Figure 3.** (a)-(d) Time series of HR-TEM images revealing the transformation of Se from a double-helix to a single-helix structure, induced by the 80 kV electron beam irradiation during observation. (a) Se double-helix contained inside a DWCNT; (b) transitional Se chain structure; (c) occurrence of a Se single-helix (red rectangle); (d) disappearance of the single-helix from the region of interest (red rectangle). Scale bar, 1 nm.

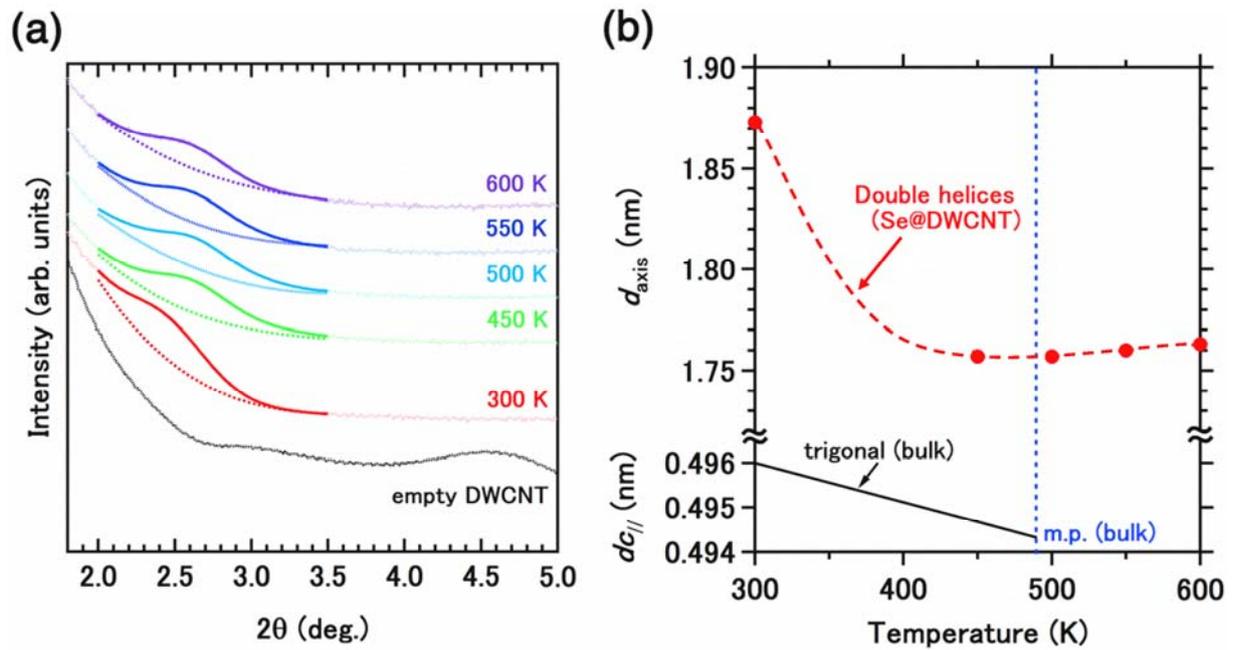

**Figure 4.** (a) XRD profiles of empty-DWCNTs and Se@DWCNTs measured at $T$=300-600 K. (b) Temperature dependence of the spatial period $d_{axis}$ in the axial direction of Se double-helices in comparison to the analogous period $d_{c//}$ along the $c$ axis of bulk Se.[1]

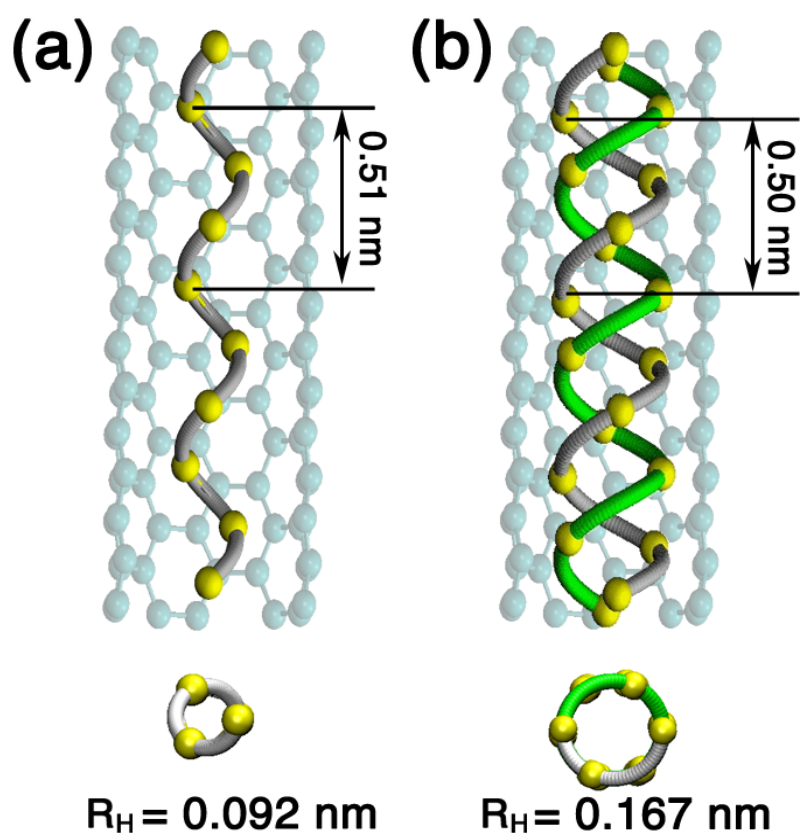

**Figure 5.** Optimized structure of a free-standing (a) single-helix and (b) double-helix structure of selenium. The faded structure of the (5,5) carbon nanotube is presented as a background for easy size comparison. The schematic connections between the Se atoms, depicting the helical structure, are shown as guides to the eye.

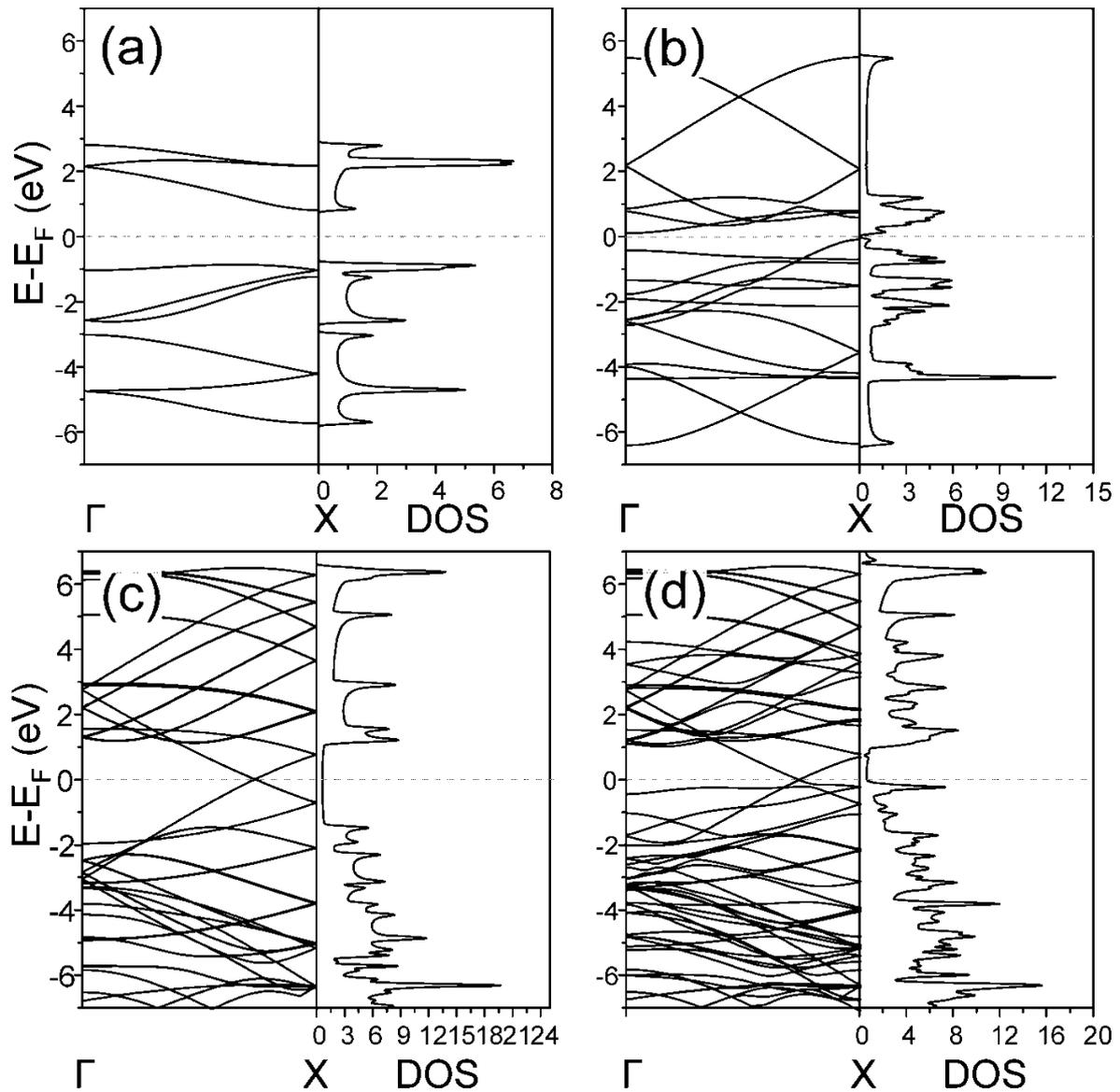

**Figure 6.** Electronic structure of a free-standing (a) single-helix and (b) double-helix of Se, (c) an isolated (5,5) carbon nanotube, and (d) a Se single-helix enclosed inside a (5,5) carbon nanotube. The left panels display the band structure and the right panels the density of states. Note the reduction of the fundamental band gap from 1.59 eV in the single-helix to 0.1 eV in the double-helix structure.

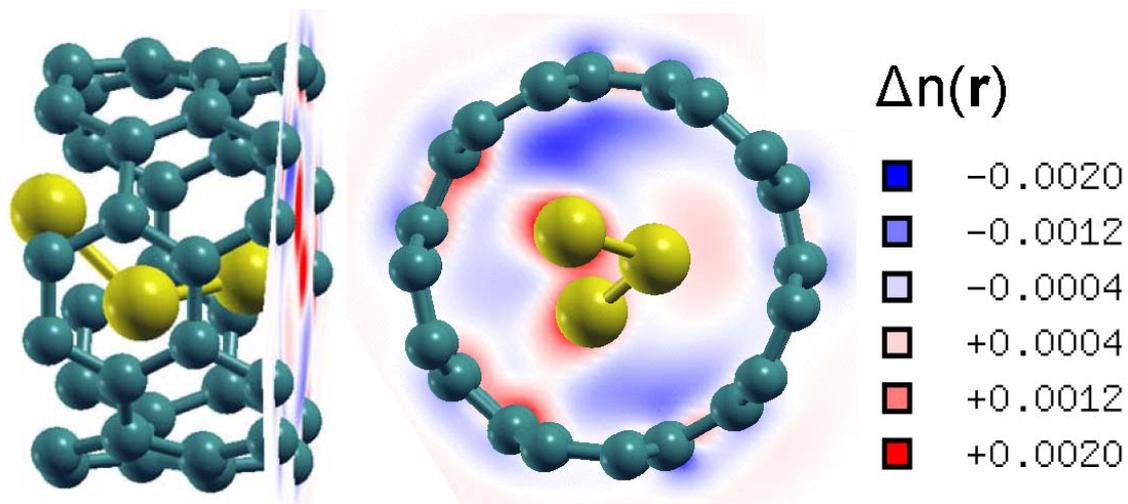

**Figure 7.** Electron density difference Δn(r) for a Se single-helix structure enclosed in a (5,5) carbon nanotube, superposed with the atomic structure. Δn(r) is defined as the difference between the total electron density of the system and the superposition of atomic charge densities and is given in el/Å$^3$ units.

ASSOCIATED CONTENT

**Supporting Information**. Details of structure optimization and simulated TEM images for uniform and nonuniform helical Se structures, vibration spectra and Mulliken charge population of Se structures in nanotubes, XPS spectra of the valence and C*1s* core level region of the Se@DWCNT system. This material is available free of charge *via* the Internet at http://pubs.acs.org.

AUTHOR INFORMATION

**Corresponding Author**

*tomanek@pa.msu.edu

**Author Contributions**

All authors have contributed to this work and have approved the final version of the manuscript.

ACKNOWLEDGMENT

T.F., M.E., and K.K acknowledge support by Exotic Nanocarbons, Japan Regional Innovation Strategy Program by the Excellence, JST. The synchrotron radiation experiments were performed at the BL02B2 of SPring-8 with the approval of the Japan Synchrotron Radiation Research Institute (JASRI) (Proposal No. 2012B1064). K.K. and T.H. were supported by Grant-in-Aid for Scientific Research (A) (No. 24241038) and (C) (No. 22510112) of the Japan Society for the Promotion of Science, respectively. RBS's visit to MSU was funded by the Sociedade Brasileira de Física and Conselho Nacional de Desenvolvimento Científico e Tecnológico (CNPq). This work was partially supported by the National Science Foundation Cooperative Agreement #EEC-0832785, titled "NSEC: Center for High-rate Nanomanufacturing".

Computational resources have been provided by the Michigan State University High Performance Computing Center. T.F. thanks Dr. K. Urita for fruitful discussion.Computational resources have been provided by the Michigan State University High Performance Computing Center. T.F. thanks Dr. K. Urita for fruitful discussion.

TOC graphic

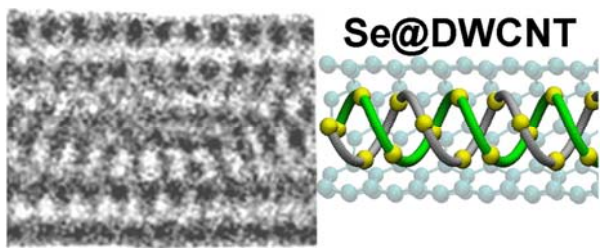